# Tracking *Peaceful Tractors* on Social Media - XAI-enabled analysis of Red Fort Riots 2021

Ajay Agarwal, Basant Agarwal

## Abstract

On 26 January 2021, India witnessed a national embarrassment from the demographic least expected from - *farmers.* People across the nation watched in horror as a pseudo-patriotic mob of farmers stormed capital Delhi and vandalized the national pride- Red Fort. Investigations that followed the event revealed the existence of a social media trail that led to the likes of such an event. Consequently, it became essential and necessary to archive this trail for social media analysis - not only to understand the bread-crumbs that are dispersed across the trail but also to visualize the role played by misinformation and fake news in this event. In this paper, we propose the *tractor2twitter* dataset which contains around 0.05 million *tweets* that were posted before, during, and after this event. Also, we benchmark our dataset with an Explainable AI ML model for classification of each *tweet* into either of the three categories - *disinformation, misinformation,* and *opinion.* These categories were obtained after a manual annotation of the entire dataset was performed by volunteer fact-checkers working in Alt News.

## RedFortRiots - Introduction

It would not be an understatement, if not an accurate comparison, to claim that the incidents that took place on 26th January 2021 in New Delhi and surrounding city borders (Tikri and Ghazipur) are a mirror of the Capital Riots of the USA - a political figure promising peaceful protest/march of thousands of protesters to a major national monument (or establishment) only for the events turning out extremely violent leading to deaths of several law-establishing authorities (i.e., police). Given, the unanticipated turn of events, a simple protest with no sight of possible violent intent ends up being a national embarrassment for the country. [1-2].

On 17th September 2020, the Lok Sabha in India passed the "Farm Bills 2020" with the simple aim of radicalization of farmers' in India by introducing privatization of *mandis*, fixing MSP (minimum support price), and other upliftment reforms. Later, on 20th September 2020, Rajya Sabha approved the bills followed by President's assent for the bills received on 27 September 2020. However, even before these legislative reforms gained the approval of the Houses, the Indian farmers led by *Samyukt Kisan Morcha* have been on pseudo-peaceful protests since 9th August 2020 in opposition to these bills. Analyzed by several economists, agricultural scientists, prior prominent figures in Indian Farmer's community, the non-Indian academic community, and other possible non-biased organizations and entities, the consensus towards these bills have been a positive one, summarizing it to be a - *"necessary and a long-sought step in the right direction to bring upliftment in the status quo of Indian farmers".* Unfortunately, due to unwanted political interventions and an overall orthodoxical approach to matters of change, the Indian farmers are sustaining their lack of evidence-backed arguments against the bills. [3].

Irrespective of their disbeliefs, provided the establishment of a democratic construct, the farmers have every constitutional right to voice their concerns - however, at the cost of constitutional rights there exists a payoff of constitutional duties primarily one being - *not interfering in practice of constitutional rights bestowed to other individuals.* At the end of the last sentence, exists a rather unfortunate realization of the dilapidation caused by misinterpretation of constitutional rights offered in a democratic establishment. [4].

Coinciding with the onset of our data collection, on 2nd January 2021, the President of BKU Sidhpur - Darshanpal announced the collective decision to hold a tractor parade in and across the Delhi contingents on 26 January (Indian Republic Day). This date was deliberately chosen to send a political and moral message regarding the role farmers play in a country is the same, if not more, than soldiers. The Delhi Police realizing the possible chaos that might ensue given the non-guided route of the Tractor Parade, police authorities along with the law-enforcing authorities held a joint conference with a delegatory body representing the farmers and jointly came up with consensually agreed routes for the execution of the Tractor Parade. Provided the vote of confidence given by the BKU and leading Farmers' Organization for following disciplinary action and parade routes during the Tractor March. [5].

However, parallel to the cluster that offered the vote of confidence to the Delhi Police, there existed a rather discrete and identifiable cluster that believed the event was a *once-in-a-lifetime* opportunity to send a rebellious and powerful message, even at the cost of public disruption and violence. This rebellious cluster, as later identified, was led by *Deep Sindhu* (prominent Punjabi actor and pretentious activist).[6]. Unforeseen and undetected by the local authorities, this rebellious cluster has strengthened its reservations on the need for an element of surprise violence to fulfill the need of sending a *powerful message to the Central authorities.*

Consequently, on an anticipated day, a series of *violations* took place leading to an uncontrolled chain of chaotic and violent turn of events. Starting with the fact that the parade started earlier than agreed upon, a group of tractors deviated from the pre-planned route (which comprised to parade around vicinities of Delhi) and entered the metropolitan locations of Delhi learning to the Red Fort. Eventually, the Capitol was replaced by the Red Forts, arms and ammunition were replaced by *loaded tractors* and horsemen, and Washington DC was replaced by New Delhi - leading to an uncomfortable and unfortunate *indigenous experience of Capital Riots in India.* The unfortunate part being that *instead of delusional and paranoid political supporters,* the indigenous experience included *nation-loving patriotic farmers with just an appeal for the Central government.*

**Data Collection**

To collect *twitter data*, we curated a list of trending keywords and hashtags. We utilized the OSINT tool *twint* to bypass the rate limits set by Twitter and collect exhaustive meta-data. [7]. This allowed to complete data collection at a faster rate and enabled archiving *twitter data* which even more than a one-week-old (as limited by Twitter official API). Henceforth, we collected *tweets* from the period 2 January 2021 to 28 February 2021. As the crisis even took place on 26 January 2021, we extended our timeframe until 28 February to record the narrative ripples that closely followed suit to the event (*AI-Face Recognition used by Delhi Police, FIR against Deep Sindhu, Greta Thunberg's toolkit, and Disha A. Ravi acquittal)* [8-10].

Approximately, 0.05 million tweets were collected. The description of our dataset has been provided in Table 1.

| *Total Tweets* | 50,351 |
|---|---|
| *Retweets* | 2,06,946 |
| *Unique Users* | 23,203 |

## 3. Data Observations

After the process of data-curation, at the onset we had four basic objectives whilst gaining insight from the data 1) *What* was being tweeted about 2) *How* polar was the content being tweeted 3) *Who* were the people tweeting and creating *engagement*, and lastly 4) How *many roles* did low-resource (non-English) language play in the narrative formation? [11].

**3.1** What was being tweeted about

To harness insight into what type of content was majority being tweeted about, we took into consideration the top 15 hashtags that were used most frequently. Normalization of the hashtag frequency was done against the number of tweets having at least one mention of a hashtag. Therefore, we computed for each top fifteen hashtags a corresponding "percentage of usage" which has been plotted in Figure 1.

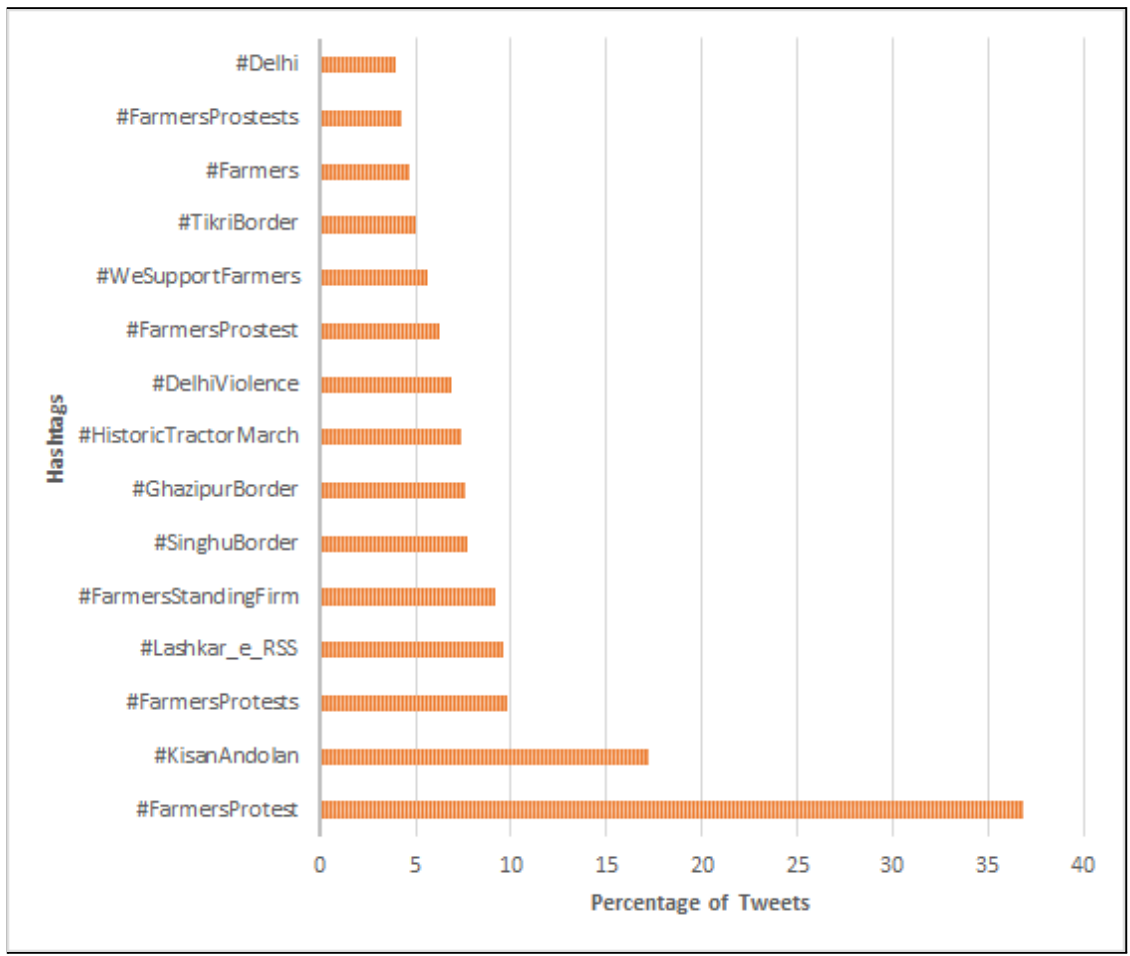

Figure 1. Top fifteen most frequently used hashtags. Apart from "#Lashkar_e_RSS" all other hashtags were usually in favor of the Tractor March without any sense of contextual polarity.

To get further insights from the collected data and analyze what content was being *tweeted* about, we plotted corresponding unigram, bigram, and trigram distributions for the top fifteen terms being utilized in the corpus of *tweets* tweeted. It must be noted that we took into account the corresponding normalization i.e. each term was normalized against the total number of *tweets* that were posted on the Twitter platform during the period of data collection. The corresponding unigram, bigram, and trigram distribution of *tweets* are shown in Figures 2, 3, and 4 respectively.

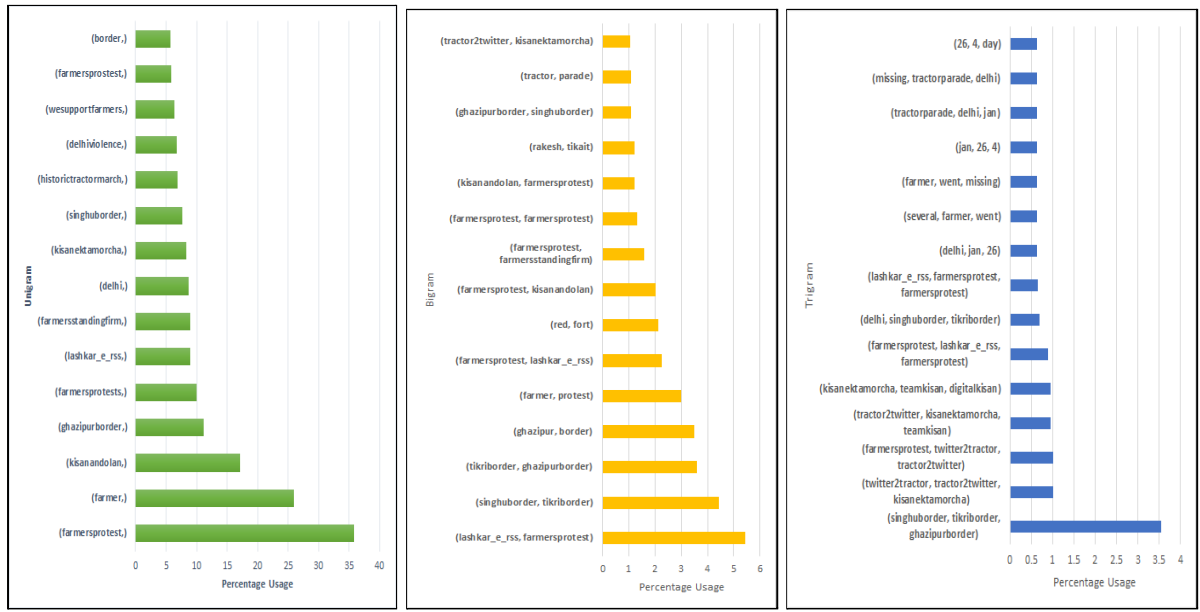

Figures 2-4. Figures 2-4 represent the top 15 unigrams, bigram, and trigram distribution normalized against the total number of *tweets* collected.

Notice from the n-gram distribution shown in the above figures. The unigram distributions follow linguistic similarity with the linguistic corpus for the top 15 hashtags used. Hence, contextual disparities are less likely to prevail i.e. the word choice for hashtags wasn't carry more than one meaning in any context. Moreover, we can also observe the few following insights from the bigram and trigram distributions -

1. The term *Lashkar_e_RSS* continues to dominate in bigram distribution and hashtag distribution highlighting that people generally carried a negative narrative of the ruling party RSS and considered the enabling of *Farm Laws* to be equivalent to terrorist activities.* However, currently such a conclusion cannot be reached as the increased usage of an extremely polar, the outlier-like term may suggest *bot activity.* This is because all other n-gram distributions suggest a general agreement with the farmer's action and narrative without being resentful or polar.
2. For the first time, we also detect opposing narratives to the RedFort riots as suggested by the unigram (delhiviolence). This means, unlike highlighted in popular media, there exists a significant public narrative that condones the activities that took place on 26 January 2021.
3. Instances of *subjectivity* are seen as more familiar names emerge in trigram and bigram distributions like - Rakesh Tikait, Tikri Border, Ghazipur Border, Red Fort, etc. Therefore, it becomes less likely that *propaganda* was at play in the social media wave.
*Lashkar_e_RSS is an amalgamation of two terms - Lashkar_e_Taiba and RSS. The former is a worldwide-spread terrorist organization responsible for carrying out terrorist activities in countries like India and the USA. The latter is the name of the ruling party in India.

**3.2** *How* polar was the content being tweeted

To understand how polar was the content that was being tweeted, we utilize the simple procedure of Sentiment Analysis using NLTK Vader. The *tweets* column was pre-processed to remove punctuations, URLs, emoji, and English stopwords (Gensim). This was followed by Sentiment Analysis done utilizing VADER. The results for the percentage of *tweets* with *positive, negative,* and *neutral* polarity have been summarized in Table 2 and visualized in Figure 5.

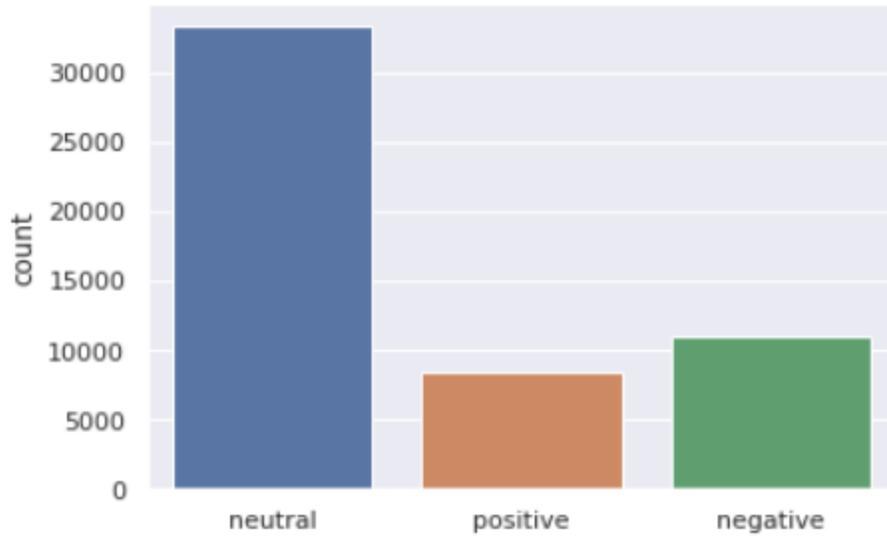

Figure 5. The above figure depicts the polarity distribution for the 50351 *tweets* in our *tractor2twitter* dataset

| Polarity Inclination | Normalized Percentage of *tweets* |
|---|---|
| *Neutral* | 66.19% |
| *Negative* | 21.9% |
| *Positive* | 16.57% |

Table 2. The above table depicts the normalized percentage of *tweets* for every polarity inclination in NLTK Vader

Notice, a majority of *tweets* were grouped under the category *neutral* which was followed by *negative* establishing the confirmation regarding the existence of wide-spread public dissent towards the Farm Bills.

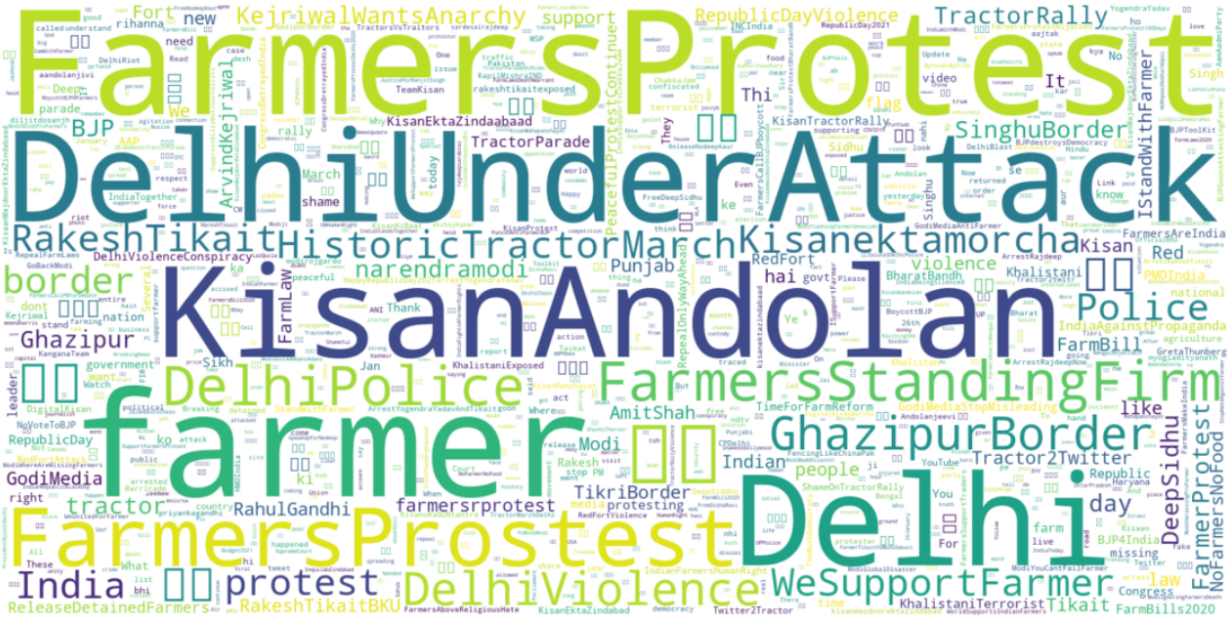

Figure 6. The above figure depicts the *wordcloud formation* for the *tweets* in the dataset. Notice, the most frequently used words include - *Kisan Andolan* which is the Hindi translation for "*Farmer Protest*"

Finally, to gather a brief overview regarding the kind of words used most frequently in this *tweet* corpus, we plot a WordCloud with pre-processed *tweets* as depicted in Figure 6.

**3.3** *Who* were the people tweeting and creating *engagement*

Provided the Twitter platform, the process of analyzing and sketching a basic user profile that fits the consensus of those who were involved in creating engagement during, before, and after the riots; we take into consideration three major parameters from our meta-data of *tractor2twitter* - mention count, post count, and retweet count.

Beginning we mention count, the top 15 accounts (Twitter handles) that were most mentioned by users in context to Red Fort Riots 2021 have been summarized in Table 3. Corresponding to each Twitter username included in Table 3, there exists the total count of mentions made to the person, along with a normalized percentage of such *mention-inclusive tweets for that person* to find the contribution of such tweets to the database.

| Username | Mention Frequency | Percentage |
| --- | --- | --- |
| rakeshtikaitbku | 458 | 0.9096 |
| youtube | 428 | 0.85 |
| delhipolice | 203 | 0.4032 |
| narendramodi | 154 | 0.3059 |
| kisanektamorcha | 143 | 0.284 |
| rahulgandhi | 124 | 0.2463 |
| isidhumoosewala | 56 | 0.1112 |
| pmoindia | 54 | 0.1072 |
| thewire_in | 47 | 0.0933 |
| bjp4india | 46 | 0.0914 |
| brajeshlive | 42 | 0.0834 |
| amitshah | 41 | 0.0814 |
| kumarkunalmedia | 40 | 0.0794 |
| incindia | 40 | 0.0794 |
| iammitesh86 | 39 | 0.0775 |

Table 3. The above table depicts the most mentioned users on Twitter along with the normalized count of mention tweets against total tweet volume.

Based on the above Table 1, there are certain parallels that we can distinctively draw with the chain of events that occurred on 26 January 2021.

1. The most mentioned username is *rakeshtikaitbku* - which belongs to the primary instigator for Red Fort Riots - Rakesh Tikait. He presides over as the head of BKU, which is an organization that stands for farmer's rights in India. Delhi Police has currently charged him with multiple felonies for the events at Red Fort.
2. Two usernames have been trending, which at first might appear as an outlier, but rather also conform with police investigations that were carried in suit with the riots - *youtube* and *isidhumoosewala*. The latter belongs to a famous Punjabi singer, whose involvement in the riots wasn't of any direct nature. However, as the investigations clarified subtle riot inciting messages were shared through subliminal undertones of comments in Youtube music videos of various famous Punjabi singers (including him). Hence, it would be rather obvious to witness his presence in the top 15 most mentioned usernames (not for riots, but as creating awareness for being a medium of communication during such riots). The author suggests that future research can be carried out to analyze the temporal volume distribution of comments in his Youtube videos during the period of riots.

With the above conclusions established, we move ahead to the second parameter for activity analysis - *retweets.* The role *retweet* plays is to depict acceptance of a particular person's opinion with the intent to share it with people and make it popular. Therefore, to analyze and identify the people whose opinion the public bought the most (at least on Twitter) during the quantum of those riots, we identified the usernames with the most cumulative retweets during the period with *tweets* that comprised of keywords from the tractor riots. This formulation for the top 15 has been depicted in Table 4.

| username | retweet_count |
| --- | --- |
| rahulgandhi | 12197 |
| kisanektamorcha | 8583 |
| amaanbali | 5090 |
| navjammu | 5019 |
| askanshul | 4702 |
| news24tvchannel | 4410 |
| tractor2twitr | 3956 |
| ifearlesssingh | 2693 |
| pragyalive | 2510 |
| boltahindustan | 2463 |
| aajtak | 2131 |
| news_18india | 2107 |
| with_kaur | 2000 |
| thecaravanindia | 1986 |
| inikhatali | 1767 |

Table 4. The above table depicts the top 15 most retweeted users on Twitter

It is interesting to note that the observations from Table 2 also conform with the events of reality that occurred during the riots. The most re-tweeted username belongs to *rahulgandhi* and *kisanektamorcha.* The former is the Twitter handle of Rahul Gandhi who is the opposition leader of the political party INC (Indian National Congress) which fills the role of opposition to the ruling party in India. It must be noted that his views (along with his party's) are highly critical of the Farm Bills (which have been considered and accepted to be the reason behind the protests leading to the riots). The latter is the Twitter handle of Kisan Ekta Morcha, which is a prominent national farmer's union opposing the Farm Bills.

Apart from these *Twitter handles* which display their public inclinations towards the bills and the riots, certain *Twitter* handles of news media outlets like AajTak, News18India, and News24 have found their way into this list due to their involvement in on-ground events during the RedFort riots. Finally, the remaining usernames occupying places in the list belong to contextually famous people with independent inclinations (of either direction) towards the people. The fact that whether these usernames belong to some individuals or some bots is a bit unclear. However, as seen with various cases, it might also be true that these might be new accounts created specifically for voicing concerns about this bill to the government and the people. Such conclusions about this potential third-group could only be substantiated positively, or negatively once an exhaustive user profiling has been done and documented.

| Username | Tweet_counts | Account Origin | Account Creation Date |
|---|---|---|---|
| jaitrejait | 527 | planet_earth | Jan 2018 |
| diljit_maan12 | 303 | USA | June 2020 |
| bebeyondthesoul | 262 | Mumbai | May 2010 |
| sikhnews247 | 236 | N/A | Aug 2020 |
| punjabtak | 229 | N/A | Jun 2018 |
| ptcnews | 193 | Mohali, India | Sept 2012 |
| jbbal75 | 191 | Calgery, Alberta | Nov 2020 |

| basu_vai | 144 | India | Jan 2018 |
| --- | --- | --- | --- |
| ansika05 | 141 | N/A | Nov 2018 |
| yespunjab | 138 | Punjab | Dec 2010 |
| saumya114 | 137 | India | Nov 2015 |
| jaspree45100801 | 129 | India | Dec 2020 |
| yadavnikhilesh1 | 127 | N/A | Apr 2020 |
| with_kaur | 126 | N/A | Feb 2021 |
| ajay75431 | 126 | India | Feb 2021 |

Table 5. The above table depicts the top 15 most active users on Twitter from January through February along with their publicly mentioned "location" and "account creation date".

Finally, we move ahead to the last straw of basic user profiling to understand the user demographic involved in Twitter activity during the riot. Our primary focus now will be - *tweet count* i.e. the number of *tweets* made by a username on the subject of riots during the period. It must be noted that this paper was written as the events unfolded post-riots which led to the origin of many significant and insignificant hypotheses emerging for further investigations. To date, while most of them have either been addressed in our paper or by police investigations, some remain open for questioning. One of the prominent being - *understanding the involvement of non-Indian origin users involvement in instigating (if not creating) an element of rebel and violence for the protests to Farm Bills, which in turn led to the unforeseen riots.*

To address this question in the most basic manner, we extended this last segment of analysis to include the publicly displayed "locations" and "account creation dates" of the top 15 most active Twitter usernames. The details of the same have been displayed in Table 3. Note, whilst the majority of usernames are of Indian origin, the fact that in a dataset of around 50,000 *tweets,* out of the top 3 usernames (and top 10) there exists one non-Indian origin username (and 2 in the top 10) - *diljit_maan12.* It must also be noticed that the second non-Indian origin username belongs from Calgery (Canada) which currently has been an active haven for the insurgent Khalistani group (as identified by NIA and Delhi Police). Instead of establishing a fact, we would rather establish a question for further analysis that involves profiling the given username for the potential pro-Khalistaani sentiment. Finally, the last observation that can be drawn from the table is that as hypothesized for any general riots, there do exists accounts created specifically for voicing concerns of Farm Bills and supporting these riots (as seen by several recent creation dates of Twitter handles).

Provided our set of observations in the last paragraph, we can; *though not conclusively;* state that due to the position of non-Indian origin username in the top 3 most active usernames during this period, a certain amount of *distinctive and non-refutable validity* has been provided the hypothesis made by Delhi Police investigations into the matter regarding the role of *international usernames* in supporting *narratives and instigating, if not creating, an element of violence and sedition in such said narratives.*

**3.4** How *much role* did low-resource (non-English) language play in the narrative formation

Post 26 January 2021, there was an upsurge for demands of investigation on the rioters and people primarily involved in instigating the event. Whilst, various discoveries were made - some backed with circumstantial evidence, while others backed with documented ones, the most prominent insight that the Delhi Police, NIA, and various other investigative authorities reached was that - the event was pre-meditated and has a non-zero scope for prior intention for disruption. This intention had been circulating through various modalities which was not just limited to even social media. Given the background established and the demographic of the people involved in the riots already known (Indian farmers), it must be questioned - *what role did the choice of language play in the riots? How did these "chosen languages" performed in contrast to their volume against English?*

To understand this, we utilized the "language" column of our *tractor2twitter* dataset which also collected metadata on the language in which the *tweet* was comprised. Note, this allocation is performed by Twitter and is often unable to identify the language used from just the text. In such cases, Twitter assigns the language code - "un" for undetermined.

Figure 7 depicts the *tweet* volume distribution for each Indian language mentioned in the *tractor2twitter* dataset. Note, a significant volume of tweets was marked as "undetermined" with 26.79%. Given an exhaustive exploratory data analysis done of the dataset, it can be determined that the *tweets* marked as undetermined (language codes) by Twitter are mostly *mixed-language coded tweets* where two or more language pairs were used to create the *tweet text* like Punjabi and Hindi.

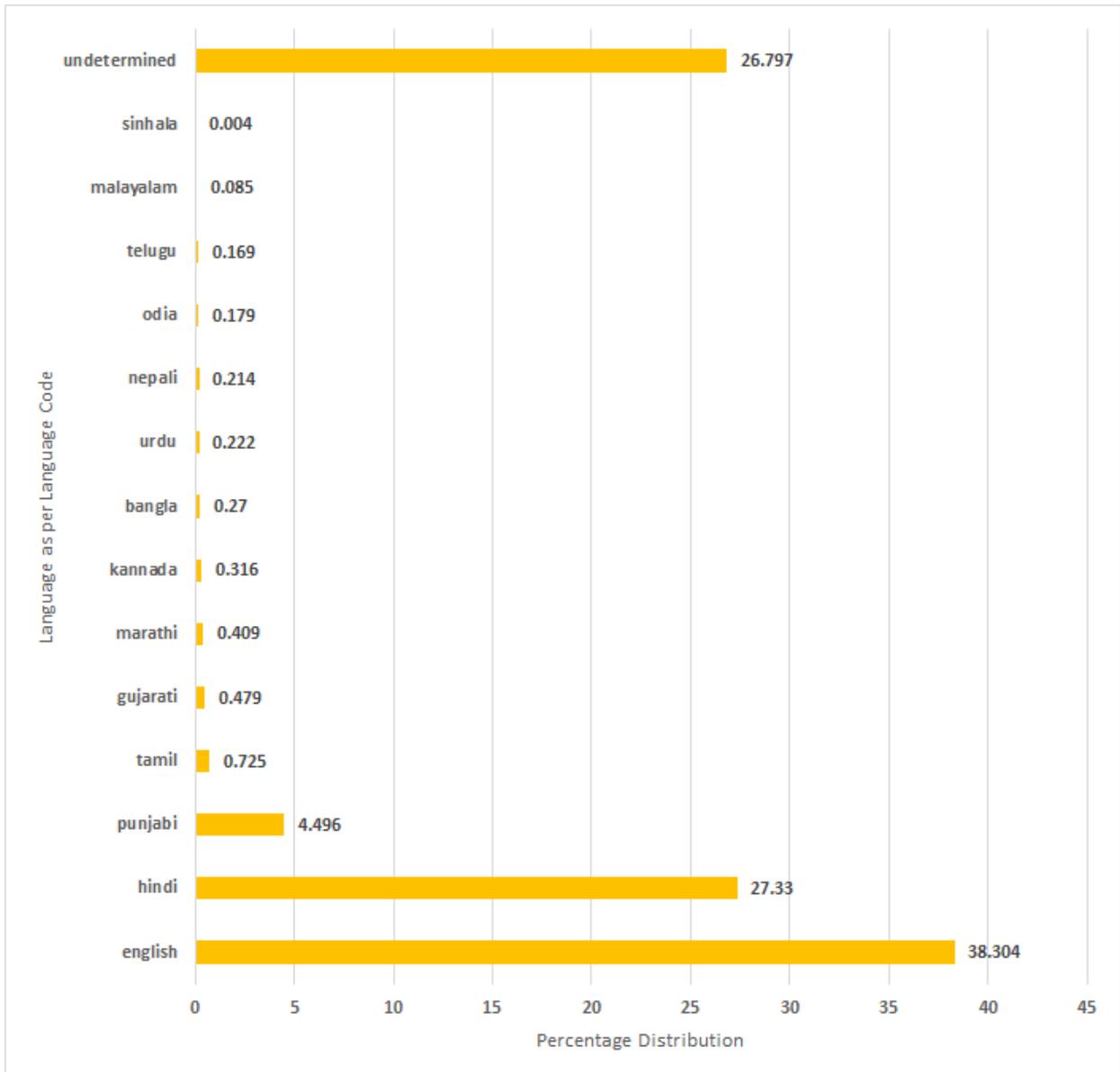

Figure 7. Percentage Distribution of *tweets* in different Indian Languages from *tractor2twitter* dataset.

For further analysis and insight generation regarding how these different languages perform in contrast to the contribution provided by English in the *tractor2twitter* dataset, we plotted a pie-chart distribution corresponding to each of the language contribution ratios with the English language as shown in Figure 8.

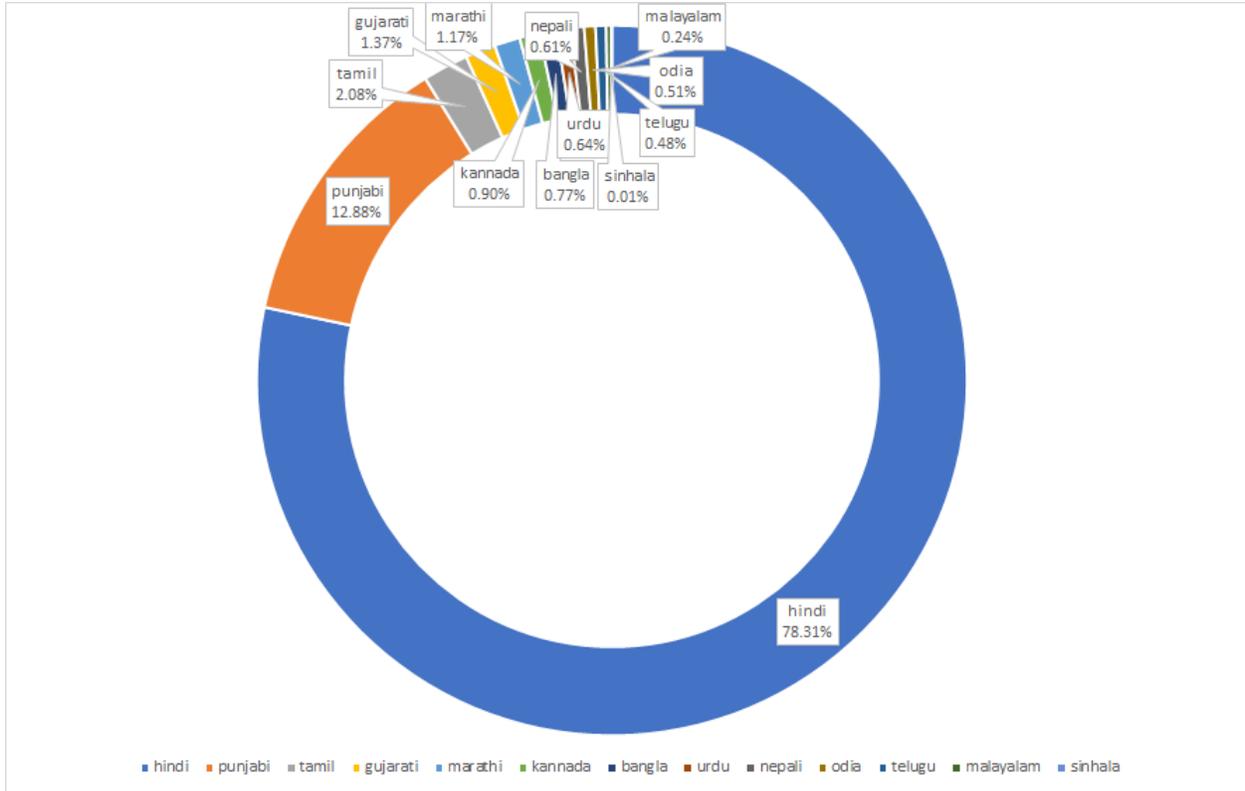

Figure 8. Individual Language contribution compared with corresponding English language contribution in the *tractor2twitter dataset*.

As seen from Figure 8, the highest competing contribution against English was provided by two languages - *Hindi* and *Punjabi* with *78.31%* and *12.88%* respectively. This also reconfirms the ground reality regarding the riots that a majority of farmers and other participants of the riots belonged to states of *Haryana* and *Punjab*. (states where Punjabi and Hindi are native languages). Also, the significant yet lesser contribution of South Indian languages like *Tamil* and *Kannada* with corresponding contributions almost equal to or above 1% indicates that narratives regarding the riots (be it positive or negative) were also voiced by South Indian farmers i.e., unlike the national assumption farmers from South were likely aware of the riots. However, this conclusion needs more decisive proofs which can be observed through user profiling of the *tractor2twitter dataset.*

## 4. Dataset Curation with Manual Annotation

The data we collected comprised around 50 thousand *tweets* with their corresponding meta-data. For utilization of the same in the domain of misinformation and disinformation research, we assign 2 randomly chosen anonymous reviewers who work at Alt News, IFCN approved fact-checking website in India. Taking sight of existing datasets of fake news spread during COVID19, and another independently conducted literature review, we come up with the following three labels - *misinformation, disinformation,* and *opinion.* Whilst the first two labels carry their accepted and defined meaning, the label *opinion* means which don't fall in either of the two categories and only represent "personal opinion of the person *tweeting*". With the manual annotation complete, we present our final dataset and name it - *tractor2twitter*. This name is based on a trigram that was identified in the top 15.

We take a train-validation-test split of 60:20:20 maintaining class-wise distribution for all three label classes. It is also interesting to note that our dataset is class-wise balanced and the same has been preserved even in the train-validation-test split. A succinct description of the dataset *tractor2twitter* has been summarized in Table 6.

| Split | Misinformation | Disinformation | Opinion | Total |
| --- | --- | --- | --- | --- |
| Training | 10575 | 10560 | 9076 | 30211 |
| Validation | 2340 | 3900 | 3830 | 10070 |
| Test | 3325 | 3425 | 3320 | 10070 |
| **Total** | 16240 | 17885 | 16226 | 50351 |

Table 6. Split-basis distribution and class-basis distribution of *tractor2twitter* dataset

## 5 Baseline and Results using ML algorithms

**Pre-processing** For pre-processing, links, hashtags, mentions, emoji, non-English words, non-alphanumeric characters, and English stopwords were removed.

**Feature Extraction** For feature extraction, we utilized tf-idf (term frequency-inverse document frequency) and Bag-of-Words

**ML Algorithms** For classification into three-labels, we utilized baseline models with Logistic Regression (LR), Support Vector Machines (SVM) with linear kernel, Decision Trees (DT), and KNN classifier (KNN) and Naive Bayes (NB). These model choices were taken based on their corresponding compatibility with local post-hoc explainability provided by various XAI libraries. Table 7 shows the accuracy, weighted average precision, weighted average recall, and weighted average F1 score for the ML models on the validation data.

| Model | Accuracy | Precision | Recall | F1 |
|---|---|---|---|---|
| LR | 87.53 | 87.56 | 87.53 | 87.50 |
| SVM | 88.78 | 88.9 | 88.78 | In ra88.56 |
| DT | 84.81 | 84.83 | 84.81 | 84.82 |
| KNN | 88.54 | 88.56 | 88.54 | 87.94 |
| **NB** | ***90.34*** | ***90.37*** | ***90.34*** | ***90.32*** |

Table 7. The above table depicts the performance metrics of various ML classifiers on validation data of the *tractor2twitter* dataset.

Based on the metrics we obtain from Table 4, we can say that the best F1 score was achieved by Naive Bayes at 90.32%. This was followed by Support Vector Machine (SVM) at 88.56%. It is also noted that the performance of Logistic Regression and KNN classifier performed closely with F1 scores of 87.50% and 87.94% respectively. In contrast, Decision Trees (DT) performed a bit inferior with an F1 score of 84.82%. With this, we benchmark our *tractor2twitter* dataset using various ML classifiers and project the same as baseline potentials.

## 6    Integrating Explainable AI for enabling Interpretability

### 6.1 Role of XAI

Given the attachment of cognitive biases in enabling susceptibility towards misinformation belief, it becomes important at the fore-front to establish our choice in the debates of *local vs global* and *self-explainable vs post-hoc*.[12-14]. To provide a succinct description of the context of these debates, we can refer to Table 8.

| local | Individual prediction explained by the AI model |
|---|---|
| global | The model's entire predictive reasoning is explained by the AI model |
| self-explainable | Capacity for reasoning and explaining is integrated into the model's capacity for prediction |
| post-hoc | Additional operations are required in the end by the model to explain its prediction |

Table 8. This table provides a very basic understanding of the terms used in the XAI domain. Combinations of these terms inquisitively determine the explainability and interpretability of the model

There exists an AI euphoria in India. It has already been established that the disproportionate assignment of aspiration towards AI has resulted in creating a self-sustaining lack of transparency which re-inforces lack of interpretability in AI model predictions. Given that the primary focus of this paper and our dataset *tractor2twitter* is to account for a systemic behavioral change in the Indian population towards self-vigilance for misinformation, it becomes important that given a lack of technical literacy in context to AI architectures, our benchmark models should emphasize explaining their predictions over their reasoning architecture. Also, as mentioned earlier the existence of AI euphoria in India has additionally resulted in the lack of accuracy of AI models at the cost of providing basic automation of tasks. Therefore, it becomes extremely crucial that our model at first must be accurate enough (not just to sustain basic automation, but also to serve its purpose) to provide a tradeoff between choice of global explainability. Hence, we conclude with our voluntary bias of *local over global* and *post-hoc over self-explainable* resulting in our choice of *local post-hoc* XAI frameworks for our benchmark ML algorithms,

## 6.2 XAI Libraries for ML benchmarks

Given our choice for *local post-hoc explainability* for enabling XAI principles in our benchmark ML algorithms, we move ahead with understanding the various explainability techniques available for providing the same in *local post-hoc*.

*Feature Importance.* This technique is heavily reliant on providing explanations of predictions by calculating the importance scores for different features that are used to output the final prediction of the AI model. Such a technique can be utilized on diverse types of features like *manual features* from feature engineering, Neural Network learned *latent features,* and even *lexical features* comprising of n-grams.

*Example-driven.* This technique works on the maxim that *"I made this prediction based on these other instances from training data"*. As stated, this technique utilizes other examples from the labeled, training data that leads to the model finding semantic similarity with the input example reaching the final prediction

*Surrogate Model.* This technique explains the prediction of the model by learning from another (more explainable) model as a proxy to the first. Being model-agnostic, i.e. this technique can be used for both choices of *global* as well as *local,* surrogate model techniques are widely used in the field of XAI-NLP.

Each of the above-mentioned explainability techniques is accompanied by a set of *operations* that it must perform to fulfill the technique and provide model-agnostic explanations and *visualization frameworks* to assure human interpretability of the provided explanations. The operations that are required by each of the above explainability techniques have been depicted in Figures 9,10,11 respectively for each technique. For our ML algorithms, we choose *surrogate models* based on two factors - First, the *surrogate model* technique requires an operation as shown in Figure 6 called *input perturbations* (described later), and second, visualization of *surrogate model* explainability can be done using *saliency highlighting* which has been documented to be most efficient in providing human interpretability with potential for initiating behavior change.

Finally, we shall describe *input perturbations* in brief detail to justify our choice for *surrogate model* explainability.

*Input perturbations* are the operation that carries the capacity of explaining a model's output for an input (say *x)* by creating random perturbations of *x* and training an explainable linear model. The state-of-the-art for this process of *input perturbations* (and *surrogate model explainability* in general*)* is LIME, which also becomes our choice for integration with our benchmark ML algorithm Naive Bayes algorithm with an F1 score of 90.32%.

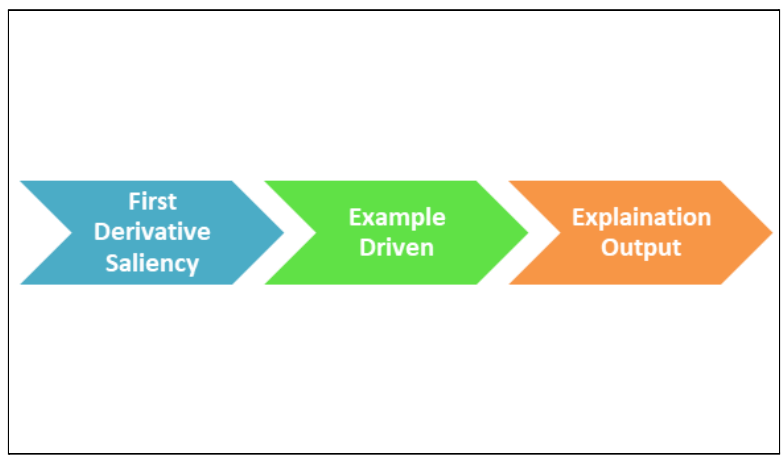

Figure 9. Explainability enabling steps required for *feature importance* explainability technique

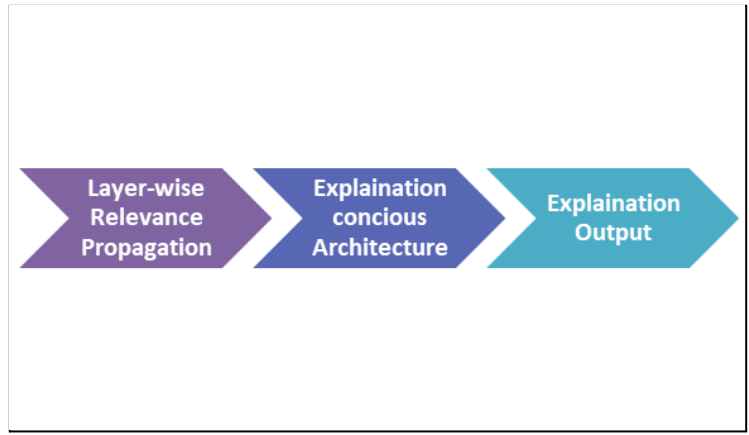

Figure 10. Explainability enabling steps required for *Example driven* explainability technique

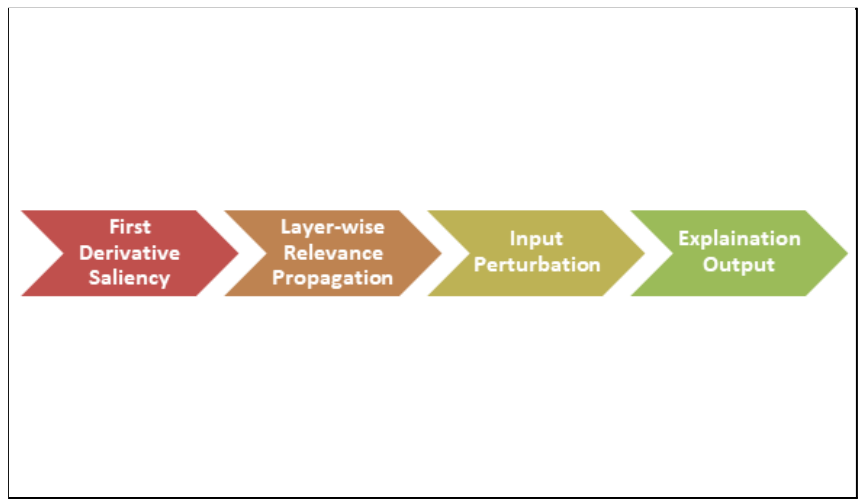

Figure 11. Explainability enabling steps required for *Surrogate model* explainability technique

### 6.3 Use-Case Demonstration

The primary end-deliverable of our XAI-enabled ML model lies in provided a deployed, web-app solution that would conform to the design we provide in Figure 12, the main focus of this section however would be to demonstrate the role played and the capacities demonstrated by the integration of the LIME library in our benchmark model (Naive Bayes). [15].

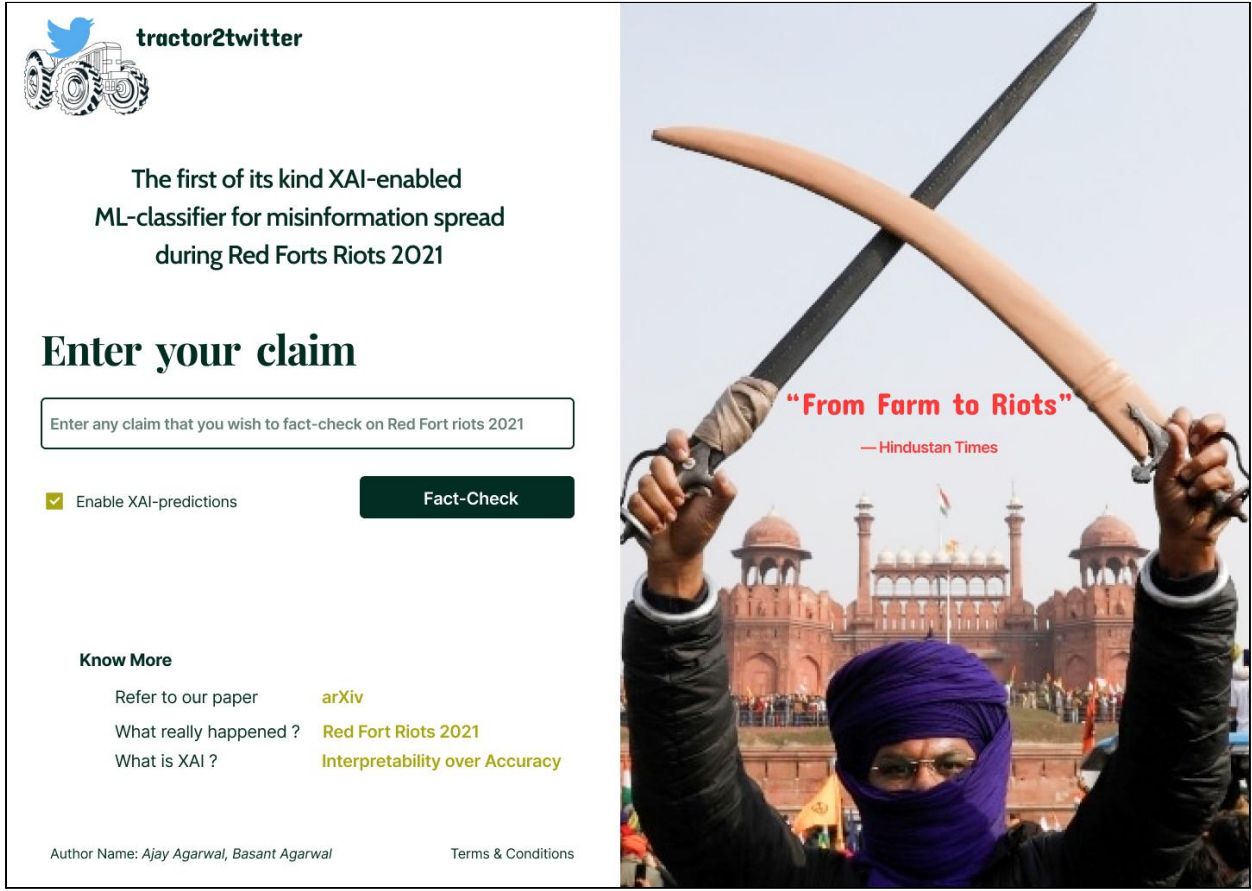

Figure 12. The above figure depicts the front-end webpage with the Naive Bayes Classifier deployed as the backend. The webpage is currently under development

As stated earlier, LIME is a technique for providing novel explanations for any ML model, considering it to be a "black box" in a human-interpretable and model-faithful manner. It utilizes the technique of *saliency highlighting* for visualizing its explanations. Being a *post-hoc local* explainability *surrogate* model, its explanations are consequences of LIME learning an interpretable model locally around the prediction of the ML classifier.

Given the integration of LIME with our benchmark model (Naive Bayes), we shall now demonstrate XAI in action by providing random claims from the test dataset. Here, LIME would perform the following three tasks following Explainable AI principles -

1. Provide visualization for probability for the given claim for each of three labels- *misinformation, disinformation,* and *opinion*
2. Provide *saliency highlighting* in the pre-processed version of the given claim to highlight which words in the claim prompted the NB classifier to assign likelihood probabilities for each label

3. Provide a visualization for the words that played the role for NB to assign probabilities to understand the probability for each word's individual likelihood to belong to a particular label.

The three claims we enter are -
1. *Red Fort riots were pre-planned by Delhi Police*
2. *Delhi Police fired bullets to curb the mob*
3. *The farmers and Delhi Police plotted the Red Fort riots together*

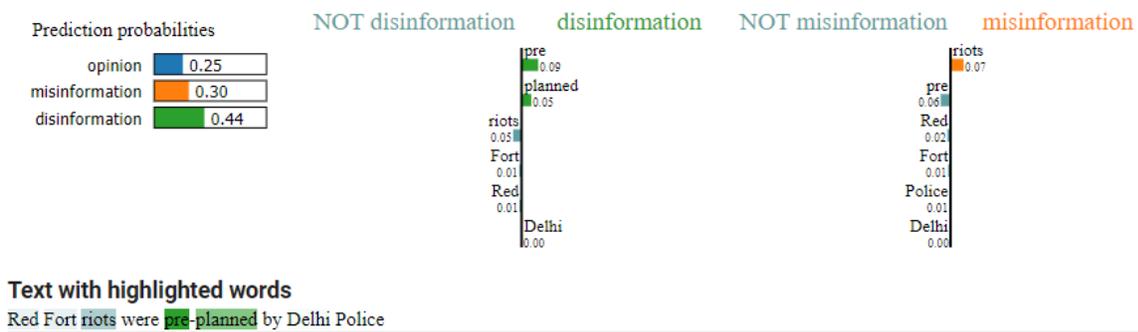

Figure 13. The above figure represents the saliency highlighting for classification for the three labels given the input claim 1. The classifier predicted *disinformation* with a probability of 44.2%. Notice, how each word that contributed to the classification has been color-coded and highlighted in the claim as well, alongside each word's probability.

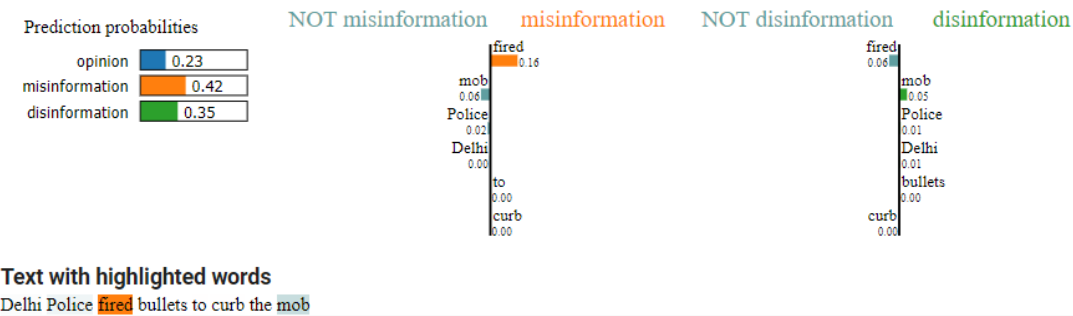

Figure 14. The above figure represents the saliency highlighting for classification for the three labels given the input claim 2. The classifier predicted misinformation with a probability of 41.95%. Notice, how each word that contributed to the classification has been color-coded and highlighted in the claim as well, alongside each word's probability.

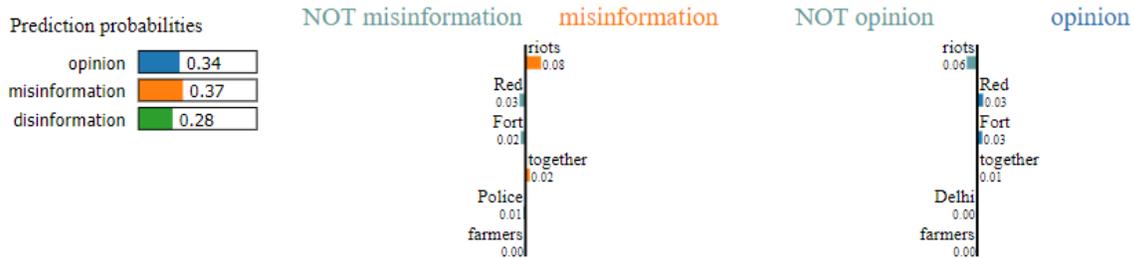

Figure 15. The above figure represents the saliency highlighting for classification for the three labels given the input claim 3. The classifier predicted misinformation with a probability of 37.11%. Notice, how each word that contributed to the classification has been color-coded and highlighted in the claim as well, alongside each word's probability.

Note, that in Figures 9-11, local probabilities for individual words have been highlighted with percentage only for the top two class labels with the highest global probabilities. Hence, for Figures 9 and 10, these two class labels are - *misinformation* and *disinformation.* For Figure 11, these two class labels are - *misinformation* and *opinion.* For global prediction probabilities for each of three random inputs provided, one can refer to Table 9.

| Input Claim | Opinion | Misinformation | Disinformation |
| --- | --- | --- | --- |
| 1 | 25.31 | 30.48 | ***44.2*** |
| 2 | 23.43 | ***41.95*** | 34.61 |
| 3 | 34.39 | ***37.11*** | 28.49 |

Table 9. The above table describes the global prediction probabilities for each of the three input samples provided to the LIME-integrated Naive Bayes (benchmark) classifier.

**Acknowledgment**

The author acknowledges and dedicates his work to his deceased grandfather and forever mentor Late Kirti Mohan Goel who had been a constant source of guidance and inspiration for the author. This work is being uploaded on the date of his unfortunate demise due to COVID19.

**Date of Upload**
24th April 2021

**Declaration of Competing Interest**

The author has no competing interest to declare

**Funding Disclosure**

The author has not been funded by any organization or institution.